\documentstyle [12pt]{article}
\begin{document}
\begin{titlepage}
\rightline{June, 1993}
\vspace{0.3cm}
\rightline{KHTP-93-03}
\rightline{SNUTP-93-21}
\vspace{0.3cm}
\centerline{\LARGE Hamiltonian formulation of $SL(3)$ Ur-KdV equation}
\vspace{1.5cm}
\centerline{\Large B.K.~Chung, K.G.~Joo, and Soonkeon~Nam\footnote{
E-mail address: nam@nms.kyunghee.ac.kr}}
\vspace{0.5cm}
\centerline {Research Institute for Basic Sciences}
\centerline {and Department of Physics,}
\centerline {~Kyung Hee University, ~Seoul, 130-701, ~Korea}
\vspace{4cm}
\centerline{\Large Abstract}
We give a unified view of
the relation between the $SL(2)$ KdV, the mKdV, and the Ur-KdV equations
through the Fr\'{e}chet derivatives and their inverses.
For this we introduce a new procedure of obtaining the Ur-KdV equation,
where we require that it has no non-local operators.
We extend this method to the $SL(3)$ KdV equation, i.e., Boussinesq(Bsq)
equation and obtain the hamiltonian structure of Ur-Bsq equation
in a simple form.
In particular, we explicitly construct the hamiltonian operator of the Ur-Bsq
system which defines the poisson structure of the system, through
the Fr\'{e}chet derivative and its inverse.
\end{titlepage}
\def\be{\begin{equation}}
\def\ee{\end{equation}}
\def\bea{\begin{eqnarray}}
\def\eea{\end{eqnarray}}
\renewcommand{\arraystretch}{1.5}
\def\ba{\begin{array}}
\def\ea{\end{array}}
\def\bce{\begin{center}}
\def\ece{\end{center}}

\def\nn{\noindent}

\def\rhu{\frac{\delta H}{\delta u}}
\def\rhv{\frac{\delta H}{\delta v}}
\def\rhq{\frac{\delta H}{\delta q}}
\def\rhj{\frac{\delta H}{\delta j}}

\def\pbv{\partial_x^5}
\def\pbth{\partial_x^3}
\def\pbs{\partial_x^2}
\def\pbt{\partial_t}
\def\pbx{\partial_x}
\def\pbi{\partial_x^{-1}}

\def\dj{{\cal D}[j]}
\def\djj{{\cal D}[j_1,j_2]}
\def\dsh{{\cal D}[\psi,\phi]}
\def\du{{\cal D}[u]}
\def\dq{{\cal D}[q]}
\def\dfu{{\cal D}^{(1)}[u]}
\def\dsu{{\cal D}^{(2)}[u]}
\def\dfuv{{\cal D}^{(1)}[u,v]}
\def\dsuv{{\cal D}^{(2)}[u,v]}
\def\dshggt{{\cal D}^{(2)}[H,G^+,G^-,T]}
\def\dfhggt{{\cal D}^{(1)}[H,G^+,G^-,T]}
\def\djjjj{{\cal D}[j_1,j_2,j_3,j_4]}

\def\fduj{\left[\frac{du}{dj}\right]}
\def\fduvjj{\left[\frac{d(u,v)}{d(j_1,j_2)}\right]}
\def\fdju{\left[\frac{dj}{du}\right]}
\def\fdjjuv{\left[\frac{d(j_1,j_2)}{d(u,v)}\right]}
\def\fdjjsh{\left[\frac{d(j_1,j_2)}{d(\psi,\phi)}\right]}
\def\fdshjj{\left[\frac{d(\psi,\phi)}{d(j_1,j_2)}\right]}
\def\fdqj{\left[\frac{dq}{dj}\right]}
\def\fdqu{\left[\frac{dq}{du}\right]}
\def\fduq{\left[\frac{du}{dq}\right]}
\def\fdhgtj{\left[\frac{d(H,G^+,G^-,T)}{d(j_1,j_2,j_3,j_4)}\right]}

\def\ho{hamiltonian operator}
Integrable models\cite{FT} in two dimensions in terms of nonlinear
differential equations have many interesting properties. In particular, they
are bi-hamiltonian systems, which means there are at least
two distinct hamiltonian structures\cite{MG}. Such a structure
provides the integrability to the system through infinite number of conserved
quantities in involution.
Another interesting feature of the integrable system is the
existence of the zero curvature condition, i.e., Lax equation which is a
compatibility condition for the pair of auxiliary linear equations.
Lax equation is crucical in solving the integrable system by the inverse
scattering method.
Recently relations between the integrable nonlinear differential equations
and two dimensional conformal field theories drew a lot of attention.
For example, the poisson brackets of
the second hamiltonian stucture of the KdV equation is the classical
version of the Virasoro algebra\cite{Ge}.
More generally, it was shown\cite{Ma} that the second hamiltonian
structure of the generalized equations of KdV type\cite{DS} is the classical
version of the extended conformal algebras, i.e., $W_N$ algebras\cite{Za}.
Furthermore Polyakov\cite{Pol} introduced a new
class of $W_N$ algebras, denoted $W_N^{(l)}$, which
was further investigated by Bershadsky\cite{Be}, and it gives the
poisson brackets of the second hamiltonian structure of the fractional KdV
equations\cite{MO,BD,Joo}. Another relation between the extended conformal
algebras $W_N$ and the generalized KdV systems is that the classical
bosonization rules for the $W_N$ algebras are given by the Miura map which is
nothing but the relation between two different choices of gauge slice\cite{BO},
and the free field representations for the $W_N^{(l)}$ algebras can be also
obtained by the same method\cite{Joo}.
\newline\indent
More recently, an action was constructed that gives the KdV or mKdV
equations as equations of motion\cite{Sc}, and their hamiltonian structures
appear as poisson bracket structures derived from this action.
Since the second hamiltonian
structure of the KdV equation is given by the local poisson brackets,
the corresponding symplectic form must contain a non-local operator,
and it follows that the action has to be non-local in the KdV variable.
However, one can write a local action\cite{Sc} for the so called ``Ur-KdV
equation", which is related to the KdV equation through the antipletic pair
formalism of Wilson\cite{Wi}.
Such a local action constitutes of two parts;
the kinetic term gives the evolution part of KdV equation that describes a
free theory in an appropriate variable, and the potential term is written in
terms of the infinite number of conserved quantities of KdV system.
\newline\indent
In this paper, we first rederive the KdV equation using zero
curvature formulation with $SL(2)$ matrix valued Lax operators,
and clarify the relation between the KdV, the mKdV, and the Ur-KdV equations
via the Fr\'{e}chet derivatives and their inverses.
In doing so, we first introduce a one parameter family of Ur-KdV equation,
and find that it becomes free of non-local pseudo-differential operators,
only for a particular value of the parameter.
We then extend this method to the $SL(3)$
KdV equation, called the Boussinesq(Bsq) equation, of which the second
hamiltonian structure is the classical $W_{3}$ algebra.
As a main result, we find the modified Bsq(mBsq) equation and
the Ur-Bsq equation which is simpler in its form than one given in
Ref.\cite{Sc} and
can be easily generalized for case of $SL(N)$ for $N>3$. We also construct the
hamiltonian operators
of mBsq and Ur-Bsq equations through the Fr\'{e}chet
derivatives and their inverse. In paticular, we see that the hamiltonian
structure of Ur-Bsq equation is given by non-local poisson brackets.
\newline\indent
The KdV equation, ${u_t=u_{xxx}+6uu_x}$,
is a bi-hamiltonian system and is obtained by zero curvature condition
parameterizing the two gauge potentials belonging to $SL(2)$ as
\be \label{1}
U=\left( \begin{array}{cc} 0 & 1\\ -u+\lambda & 0 \end{array}\right) , \quad
V=\left( \begin{array}{cc} A(x,t) & B(x,t)\\ C(x,t) & -A(x,t)\end{array}
\right) \, ,
\ee
with a constant spectral parameter $\lambda$, and the corresponding Lax
operator is $L=\pbx-U$. The zero curvature condition
 $\left[\pbx -U, \pbt -V\right]=0$
would lead to the dynamical equation and the constraint equations. Putting
$B={\displaystyle \rhu}$,
the dynamical equation can be expressed as a hamiltonian system:
\be \label{2}
u_t=\left[\dsu -2\lambda{\cal D}^{(1)}(x)\right]\rhu \, ,
\ee
where the first hamiltonian operator, ${\cal D}^{(1)}(x)$,
and the second hamiltonian operator, ${\dsu}$, are given as
\be \label{3}
{\cal D}^{(1)}[u]=\pbx , \quad	{\dsu}=\frac{1}{2}{\pbth} + 2u\pbx+u_x \, .
\ee
This equation expresses the feature encountered in the hamiltonian analysis of
integrable hierachies, i.e., the presence of two coordinated poisson
structures expressed as the one-parameter family of brackets;
$\left\{u(x), u(y)\right\}=\left\{u(x), u(y)\right\}_1 + \mu \left\{u(x),
u(y)\right\}_2$ ($\mu$ is arbitrary).
Taking $B={\displaystyle  \rhu} =2u$, the KdV equation
is given as the second hamiltonian equation,
$u_t={\dsu}{\displaystyle \rhu}$, and the corresponding poisson bracket is
$\left\{u(x), u(y)\right\}= \dsu \delta(x-y)$.
This poisson structure represents the classical part of the Virasoro
algebra\cite{Ge}.
\newline\indent
To obtain the standard Miura map, we perform the gauge
transformation $L\longrightarrow\Phi^{-1}L \Phi$ for the Lax operator $L$ with
$\Phi(x,t)$ taking values in the strictly lower triangular matrices of $SL(2)$
with the diagonal elements set equal to one. Setting the transformed Lax
operator as
$\tilde{L}=\pbx-\tilde{U}$, where $\tilde{U}=\Phi^{-1}U \Phi-\Phi^{-1}\Phi_x$,
and choosing a gauge as $\tilde{U}_{11}=-j, \tilde{U}_{22}=j,
\tilde{U}_{12}=1$,
and $\tilde{U}_{21}=\lambda$,  we obtain the well-known Miura map;
$u=j_x - j^2$. For the evolution equation of the variable $j$, that is
modified
KdV equation(mKdV), we introduce the Fr\'{e}chet derivative of $u$ with
respect to $j$, ${\displaystyle \fduj=\pbx-2j}$.
In general, given a transformation $u=F[j, j_{x}, \cdots ]$,
${\displaystyle \fduj}$ is
the differential operater that implies $u_t={\displaystyle \fduj} j_t$, and
for any functional $H[u]$, we have
${\displaystyle \rhj}={\displaystyle {\fduj}^{\ast} \rhu}$, where
${\displaystyle {\fduj}^{\ast}}$
is the formal adjoint of ${\displaystyle \fduj}$.
The hamiltonian operator of mKdV equation, ${\dj}$, which defines
its hamiltonian equation, ${\displaystyle j_t=\dj \rhj}$, is related to the
second hamiltoian operator of KdV equation through the above definition of
the Fr\'{e}chet derivative and their formal adjoint\cite{Wi}, as follows:
\be \label{4}
{\dsu}=\fduj \dj {\fduj}^{\ast}\, .
\ee
In the above, we have ${\displaystyle {\fduj}^{\ast}}=-\pbx-2j$, and it
is easy to show that $\dj=-\frac{1}{2}\pbx$ satisfies this equation.
Therefore we can easily obtain the mKdV equation as follows:
\be \label{4-1}
j_t=\dj{\fduj}^{\ast} \rhu = j_{xxx}-6j^2j_x \, ,
\ee
where we used ${\displaystyle\rhu=2u}$ of the KdV equation.
\newline\indent
Now we note that there should be an inverse operator of
${\displaystyle \fduj}$ in order to define ${\dj}$ in eq.(\ref{4}).
This suggests that we must take the form of transformation with
respect to $j$ as
$j=a\rho_{x}\rho^{-1}$ for the factorization of the Fr\'{e}chet
derivative ${\displaystyle \fduj}$, where $a$ is a constant.
If we reset $\rho=q_{x}$, we have a transformation called the Cole-Hopf map;
${j=a q_{xx}q_{x}^{-1}}$.
Using this transformation we can factorize the Fr\'{e}chet derivative and
their formal adjoint, in terms of variable $q$,
as ${\displaystyle \fduj}=q_x^{2a}\pbx q_{x}^{-2a}$,
${\displaystyle {\fduj}^{\ast}}=-q_{x}^{-2a} \pbx q_{x}^{2a}$, respectively,
where we will be using pseudo-differential operator, $\pbi$,
which satisfies $\pbx\pbi f=\pbi\pbx f=f$.
Therefore the inverses of these differential operators are written by
\be \label{5}
{\fduj}^{-1}=\fdju=q_x^{2a} \pbi q_{x}^{-2a}, \quad
\left({\fduj}^{\ast}\right)^{-1}={\fdju}^{\ast}=-q_{x}^{-2a}
\pbi q_{x}^{2a}\, .
\ee
Through this Cole-Hopf map and the Miura map, we
have also a relation between the variables $u$ and $q$ as follows:
\be\label{6}
u=a\left\{q_{xxx}q_{x}^{-1}-(a+1)q_{xx}^{2}q_{x}^{-2}\right\}\, .
\ee
Using this relation, the second hamiltonian operator of KdV system
can also be expressed in terms of variable $q$;
$\dsu=\frac{1}{2}q_x^{2a}\pbx q_{x}^{-2a}\pbx
q_{x}^{-2a}\pbx q_x^{2a}$.
{}From this observation, we can easily find the hamiltonian operator of mKdV
equation.
\newline\indent
Now we can obtain the evolution equation of variable $q$,
through the Fr\'{e}chet derivative of $j$ with respect to $q$,
${\displaystyle\fdqj}=\frac{1}{a}\pbi q_x \pbi$, and the mKdV equation, as
follows:
\be\label{7}
q_t=\fdqj j_t=q_{xxx} -\pbi \left\{
3q_{xxx}q_{xx}q_{x}^{-1}+2(a^2-1)q_{xx}^3q_x^{-2}\right\} \, .
\ee
Taking $a^2=\frac{1}{4}$, we can eliminate the pseudo-differential operator,
$\pbi$, which is not adequate in an evolution equation. Then the above
equation is the Ur-KdV equation given
in Ref.\cite{Wi}. The constant factor $a$ of the Cole-Hopf map can take two
values, $\frac{1}{2}, -\frac{1}{2}$, but two choices are not independent;
they are related by setting $j$ to $-j$.
When we take $a=\frac{1}{2}$, the transformation
eq.(\ref{6}) is the well-known Schwartzian derivative of $q$ to $2u$.
Using the Fr\'{e}chet
derivative of $j$ with respect to $q$ (or $u$ to $q$) and their inverses,
we can find the hamiltonian operator, ${\dq}$, which defines
the hamiltonian equation of Ur-KdV equation, as follows:
\be \label{8}
\dq=\fdqj \dj {\fdqj}^{\ast}=\fdqu\du{\fdqu}^{\ast}=-2\pbi q_x\pbi q_x\pbi\, ,
\ee
where,
\begin{eqnarray*}
& &\fdqj=2\pbi q_x \pbi , \quad {\fdqj}^{\ast}=2\pbi q_x \pbi \, ,\\
& &\fdqu=2\pbi q_x \pbi q_x \pbi q_{x}^{-1}, \quad {\fdqu}^{\ast}=-2q_{x}^{-1}
\pbi q_x \pbi q_{x}\pbi \, ,
\end{eqnarray*}
and we have taken $a=\frac{1}{2}$.
The non-local poisson brackets of the Ur-KdV equation are given by
the hamiltonian operator as follows: $\left\{q(x),
q(y)\right\}$=$-2\pbi
q_x \pbi q_x \pbi\delta(x-y)$. Finally we can see the exact relation
between three hamiltonian operator ${\dsu, \dj}$ and
$\dq$ using the Fr\'{e}chet derivatives and their inverses.
\newline\indent
As a generalization of the above procedure, we can find the $SL(3)$ Ur-KdV
equation by requiring that it has no non-local operators, and also find its
hamiltonian structure.
The two gauge potentials belonging to $SL(3)$ is
parametrized as
\be\label{10}
U=\left( \begin{array}{ccc} 0 & 1 & 0 \\ 0 & 0 & 1 \\ u_1+\lambda & u_2 &
0\end{array}\right),\quad
V=\left( \begin{array}{ccc} A(x,t) & B(x,t)  & C(x,t)  \\
D(x,t)	& E(x,t)  & F(x,t)  \\ G(x,t)  & H(x,t)  & I(x,t)
\end{array}\right), \quad  Tr\ V=0 \, ,
\ee
with the constant parameter $\lambda$, and the corresponding Lax
operator is $L=\partial_x-U$. Using
the zero curvature condition $[\partial_x-U, \partial_t-V]=0$, we obtain the
dynamical equations of $u_1$, $u_2$ and the constraint equations\cite{DHR}.
These equations are expressed only in terms of the variables of upper
triangular part of $V$, i.e. $B$, $C$ and $F$.
Redefining  $u=u_2, \quad v=u_{2x}-2u_1$, we recover the standard
expression\cite{MO,BD}. Using one of the constraint equations,
$C_{x}+B-F=0$, and setting $\tilde{F}=F-\frac{1}{2}C_x$,
$\tilde{C}=-\frac{1}{2}C$, we can write the hamiltonian equation of $u$, $v$
variables in a matrix form:
\be \label{11}
\left( u_t ,\quad  v_t \right)^{\dagger}=\left(\dsuv-6\lambda
\dfuv\right)\left( \tilde{F} ,\quad  \tilde{C} \right)^{\dagger} \, ,
\ee
where the subscript $t$ denotes the time derivative.
Only non-vanishing component of the first hamiltonian operator, $\dfuv$,
are ${{\cal D}^{(1)}}_{12}={{\cal D}^{(1)}}_{21}=\pbx$.
The second hamiltonian operator, $\dsuv$, is given as follows:
\be \label{12}
\dsuv=
\left( \begin{array}{cc} -2\pbth+2u\pbx + u_x & 3v\pbx +2v_x \\ 3v\pbx +
v_x  & {{\displaystyle 2\pbv /3 -10u\pbth /3 - 5u_x \pbs} \atop
{\displaystyle +(8u^2/3 -3u_{xx})\pbx -2u_{xxx}/3 + 8uu_{x}/3}}
\end{array}\right)\, .
\ee
When we take the upper triangular part of $V$ as the dual of the upper
triangular part of $U$, i.e. $B=F=0$, $C=1$ ($\tilde{F}=0$,
$\tilde{C}=-\frac{1}{2}$), we obtain the Bsq equation,
$u_t=-v_x, \quad v_t=\frac{1}{3}u_{xxx} - \frac{4}{3}uu_x$.
The other elements of $V$ are obtained in terms of $u$, $v$
variables by the constraint equation.
The above hamiltonian operators define the poisson structure
of Bsq equation.
The poisson brackets represented by the second hamiltonian operator
are the classical version of $W_3$ algebra\cite{Ma}.
\newline\indent
To look for the Miura map of the Bsq equation, that is the free field
representation of the classical $W_3$ algebra, we perform the gauge
transformation $L\longrightarrow\Phi^{-1}L \Phi$ for the Lax operator $L$ with
$\Phi(x,t)$ taking values in the strictly lower triangular matrices of $SL(3)$
with the diagonal elements set equal to one. Setting the transformed Lax
operator as
$\tilde{L}=\pbx-\tilde{U}$, where $\tilde{U}=\Phi^{-1}U \Phi-\Phi^{-1}\Phi_x$,
and choosing a gauge as
\be \label{13}
\tilde{U}=
\left( \begin{array}{ccc} j_1 & 1 & 0  \\ 0 & -(j_1 + j_2) & 1 \\ \lambda
& 0 & j_2 \end{array}\right) ,
\ee
we obtain the Miura map of $u$ and $v$ with respect to $j_1$ and $j_2$,
\bea \label{14}
u\!\!\!\!&=&\!\!\!\!(j_{1}-j_{2})_x + j_{1}^{2} + j_{1}j_{2} +
j_{2}^{2} \, , \nonumber\\
v\!\!\!\!&=&\!\!\!\!-(j_1 + j_2)_{xx} -(2j_1 - j_2)j_{1x} + (2j_2 - j_1)j_{2x}
+ 2(j_1
+ j_2)j_{1}j_{2} \, .
\eea
The modified Bsq equation(mBsq) for $j_1$, $j_2$ and its hamiltonian structure
can be exhibited by the Fr\'{e}chet derivative and their formal adjoint of the
above transformation, which is, in a matrix form,
\bea \label{15}
\fduvjj\!\!\!\!&=&\!\!\!\!\left( \begin{array}{cc} \pbx+2j_1+j_2 &
-\pbx+2j_2+j_1 \\
{{\displaystyle -\pbs+(j_2-2j_1)\pbx-2j_{1x}}\atop
{\displaystyle -j_{2x}+2j_{2}^{2}+4j_{1}j_{2}}}
& {{\displaystyle -\pbs+(2j_2-j_1)\pbx+j_{1x}}\atop
{\displaystyle+2j_{2x}+4j_{1}j_{2}+2j_{1}^{2}}}
\end{array}\right) \, , \nonumber \\
{\fduvjj}^{\dagger}\!\!\!\!&=&\!\!\!\!\left( \begin{array}{cc} -\pbx+2j_1+j_2 &
{-\pbs-(j_2-2j_1)\pbx -j_{2x}+2j_{2}^{2}+4j_{1}j_{2}} \\ \pbx+2j_2+j_1 &
{-\pbs-(2j_2-j_1)\pbx+2j_{1x}+4j_{1}j_{2}+2j_{1}^{2}}
\end{array}\right)\, .
\eea
The hamiltonian operator of mBsq equation, $\djj$, which defines
its hamiltonian equation,
and the second hamiltonian operator of Bsq equation, $\dsuv$, are related by,
through the Fr\'{e}chet derivative and their formal adjoint,
\be \label{16}
\dsuv=\fduvjj \djj {\fduvjj}^{\dagger} \, .
\ee
Now we come to our main point. As mentioned before, we need to have a
transformation by which we can
obtain the inverses of the Fr\'{e}chet derivative and their formal adjoint,
eq.(\ref{15}), in order to find $\djj$. Therefore we write the
transformation which is the generalized Cole-Hopf map as follows:
\be \label{17}
j_1=\alpha\psi_{xx}{\psi}_{x}^{-1}+\beta{\phi}_{xx}{\phi}_{x}^{-1} , \quad
j_2=\gamma{\psi}_{xx}{\psi}_{x}^{-1}+\delta{\phi}_{xx}{\phi}_{x}^{-1} \, ,
\ee
where, $\alpha,\beta,\gamma$ and $\delta$ are abitrary constants.
We see that this form of Cole-Hopf map can be easily generalized to the
$SL(N)$, $N>3$ case. Since the calculation involving arbitrary values of
$\alpha,\beta,\gamma,\delta$ is quite lengthy, we sketch the steps here.
It turns out that only particular values of $\alpha,\beta,\gamma,\delta$,
are picked out. We later go through the steps more explicitly with a
particular solution of $\alpha,\beta,\gamma,\delta$.
Using this map, we can factorize the Fr\'{e}chet derivatives, eq.(\ref{15}),
and find their inverses, denoted $\fdjjuv, \fdjjuv^{\dagger}$.
After a lengthy calculation,
the hamiltonian operator of mBsq equation, $\djj$, is obtained by
eq.(\ref{16}),
through expression of the second hamiltonian operator of Bsq equation in terms
of $\psi, \phi$. Then, it is easy to obtain the mBsq equation.
The hamiltonian operator of Ur-Bsq equation(the evolution equation of
variables $\psi, \phi$), $\dsh$, is related by the hamiltonian operator of
mBsq equation, through the Fr\'{e}chet derivatives of the above transformation
as the eq.(\ref{16}). It is not difficult to find the inverse Fr\'{e}chet
derivatives of $j_1, j_2$ with respect to $\psi, \phi$ and its formal adjoint,
$\fdshjj, {\fdshjj}^{\dagger}$, and we can obtain the non-local hamiltonian
operator of Ur-Bsq equation. Finally, we can find the Ur-Bsq equation
through the mBsq equation as follows:
\bea \label{17-1}
\psi_t\!\!\!\!&=&\!\!\!\!a\psi_{xx}+\pbi(b\psi_{x}\phi_{xxx}\phi_x^{-1}
+c\psi_{xx}^2\psi_x^{-1}+d\psi_x\phi_{xx}^2\phi_x^{-2}
+e\psi_{xx}\phi_{xx}\phi_x^{-1}) \, , \nonumber \\
\phi_t\!\!\!\!&=&\!\!\!\!f\phi_{xx}+\pbi(g\phi_{x}\psi_{xxx}\psi_x^{-1}
+h\phi_{xx}^2\phi_x^{-1}+k\phi_x\psi_{xx}^2\psi_x^{-2}
+l\phi_{xx}\psi_{xx}\psi_x^{-1}) \, ,
\eea
where,
\bea \label{17-2}
a\!\!\!\!&=&\!\!\!\!\beta(2\alpha+\gamma)+\delta(\alpha+2\gamma),\quad
b=2(\beta^2+\beta\delta+\delta^2),\nonumber \\
c\!\!\!\!&=&\!\!\!\!\beta(2\alpha^2-2\alpha+2\alpha\gamma-\gamma-\gamma^2)+\delta(\alpha^2-
\alpha-2\alpha\gamma-2\gamma-2\gamma^2),\nonumber \\
d\!\!\!\!&=&\!\!\!\!\beta(2\beta^2-2\beta+2\beta\delta-\delta-\delta^2)+\delta(\beta^2-\beta-2
\beta\delta-2\delta-2\delta^2), \nonumber \\
e\!\!\!\!&=&\!\!\!\!2(2\alpha\beta\delta-2\beta\gamma\delta+2\alpha\beta^2-2\gamma\delta^2-
\alpha\delta^2+\beta^2\gamma),\nonumber \\
f\!\!\!\!&=&\!\!\!\!-\alpha(2\beta+\delta)-\gamma(\beta+2\delta),\quad
g=-2(\alpha^2+\alpha\gamma+\gamma^2), \\
h\!\!\!\!&=&\!\!\!\!\alpha(-2\beta^2+2\beta-2\beta\delta+\delta+\delta^2)-\gamma(\beta^2-\beta-
2 \beta\delta-2\delta-2\delta^2), \nonumber \\
k\!\!\!\!&=&\!\!\!\!\alpha(-2\alpha^2+2\alpha-2\alpha\gamma+\gamma+\gamma^2)
-\gamma(\alpha^2- \alpha-2\alpha\gamma-2\gamma-2\gamma^2), \nonumber \\
l\!\!\!\!&=&\!\!\!\!2(-2\alpha\beta\gamma+2\alpha\gamma\delta-2\alpha^2\beta+2
\gamma^2\delta- \alpha^2\delta+\beta\gamma^2).\nonumber
\eea
These evolution equations include the pseudo-differential operator, $\pbi$,
which is not physically adequate in evolution equation. This can be avoided
if we require following conditions for $b=-d=e, c=0$ and $g=-k=l, h=0$.
There are six nontrivial solution sets of $(\alpha, \beta, \gamma,\delta)$
which satisfy these conditions. They are
$(\alpha, \beta, \gamma, \delta)$=$(-\frac{2}{3}, -\frac{1}{3}, \frac{1}{3},
\frac{2}{3})$,
$(-\frac{1}{3}, \frac{1}{3}, -\frac{1}{3}, -\frac{2}{3})$,
$(-\frac{1}{3}, -\frac{2}{3}, \frac{2}{3}, \frac{1}{3})$,
$(\frac{1}{3}, \frac{2}{3}, \frac{1}{3}, -\frac{1}{3})$,
$(\frac{1}{3}, -\frac{1}{3}, -\frac{2}{3}, -\frac{1}{3})$,
$(\frac{2}{3}, \frac{1}{3}, -\frac{1}{3}, \frac{1}{3})$.
In fact, these six sets do not give  independent Cole-Hopf map of
eq.(\ref{17}),
but they are related to each other through the adequate linear
combination of $j_1, j_2$.
The Ur-Bsq equations and its hamiltonian operators given by six solutions
are different from only constant factors respectively.
We will pick up one of the six solution sets, i.e.,
$(\alpha, \beta, \gamma, \delta)$=
$(-\frac{2}{3}, -\frac{1}{3}, \frac{1}{3}, \frac{2}{3})$,
and derive the above-sketched steps.
Firstly, using the transformation eq.(\ref{17}) the elements of the
Fr\'{e}chet
derivatives eq.(\ref{15}) are factorized in terms of $\psi, \phi$ as
\bea \label{18}
\fduvjj\!\!\!\!&=&\!\!\!\!\left( \begin{array}{cc}
{\psi}_{x}\pbx{\psi}_{x}^{-1} &
-{\phi}_{x}\pbx{\phi}_{x}^{-1} \\
-{\psi}_{x}^{\frac{2}{3}}{\phi}_{x}^{\frac{4}{3}}\pbx{\psi}_{x}^{\frac{1}{3}}
{\phi}_{x}^{-\frac{4}{3}}\pbx{\psi}_{x}^{-1} &
-{\phi}_{x}^{\frac{2}{3}}{\psi}_{x}^{\frac{4}{3}}\pbx{\phi}_{x}^{\frac{1}{3}}
{\psi}_{x}^{-\frac{4}{3}}\pbx{\phi}_{x}^{-1}
\end{array}\right) \, , \nonumber \\
{\fduvjj}^{\dagger}\!\!\!\!&=&\!\!\!\!\left( \begin{array}{cc}
-{\psi}_{x}^{-1}\pbx{\psi}_{x} &
-{\psi}_{x}^{-1}\pbx{\phi}_{x}^{-\frac{4}{3}}{\psi}_{x}^{\frac{1}{3}}\pbx{\psi
}_{x}^{\frac{2}{3}}{\phi}_{x}^{\frac{4}{3}}
\\ {\phi}_{x}^{-1}\pbx{\phi}_{x} &
-{\phi}_{x}^{-1}\pbx{\phi}_{x}^{\frac{1}{3}}{\psi}_{x}^{-\frac{4}{3}}
\pbx{\phi}_{x}^{\frac{2}{3} } {\psi}_{x}^{\frac{4}{3}}
\end{array}\right) \, .
\eea
By the definition of the Fr\'{e}chet derivatives and their formal adjoint,
the inverse of these differential operator are
\bea\label{19}
\fdjjuv \!\!\!\!&=&\!\!\!\!\frac{1}{2} \left( \begin{array}{cc}
{\psi}_{x}\pbi{\phi}_{x}\pbi{\phi}_{x}^{-\frac{1}{3}}{\psi}_{x}^{\frac{1}{3}}\pbx{\phi}_{x}^{-\frac{2}{3}}
{\psi}_{x}^{-\frac{4}{3}} &
-{\psi}_{x}\pbi{\phi}_{x}\pbi{\phi}_{x}^{-1}{\psi}_{x}^{-1}
\\
-{\phi}_{x}\pbi{\psi}_{x}\pbi{\phi}_{x}^{\frac{1}{3}}{\psi}_{x}^{-\frac{1}{3}}
\pbx{\phi}_{x}^{-\frac{4}{3}}{\psi}_{x}^{-\frac{2}{3}} &
-{\phi}_{x}\pbi{\psi}_{x}\pbi{\psi}_{x}^{-1}{\phi}_{x}^{-1}
\end{array}\right)\, , \nonumber \\
{\fdjjuv}^{\dagger}\!\!\!\!&=&\!\!\!\!\frac{1}{2}\left(\begin{array}{cc}
-{\psi}_{x}^{-\frac{4}{3}}{\phi}_{x}^{-\frac{2}{3}}\pbx{\psi}_{x}^{\frac{1}{3}
}{\phi}_{x}^{-\frac{1}{3}}\pbi{\phi}_{x}\pbi{\psi}_{x} &
{\psi}_{x}^{-\frac{2}{3}}{\phi}_{x}^{-\frac{4}{3}}\pbx{\phi}_{x}^{\frac{1}{3}}
{\psi}_{x}^{-\frac{1}{3}}\pbi{\psi}_{x}\pbi{\phi}_{x}
\\ -{\phi}_{x}^{-1}{\psi}_{x}^{-1}\pbi{\phi}_{x}\pbi{\psi}_{x} &
-{\phi}_{x}^{-1}{\psi}_{x}^{-1}\pbi{\psi}_{x}\pbi{\phi}_{x}
\end{array}\right)\, .
\eea
Using the above operators and expressing the second hamiltonian operator of
Bsq equation in terms of $\psi, \phi$ through the transformation,
eq.(\ref{17}),
we can find the hamiltonian operator of mBsq equation through eq.(\ref{16}),
as follows:
\be \label{20}
\djj =\frac{1}{3}\left( \begin{array}{cc} 2 & -1 \\
-1 & 2 \end{array}\right)\pbx \, .
\ee
Then the mBsq equation\cite{Sc} is
\be \label{21}
{\left(\begin{array}{c} j_{1t} \\ j_{2t} \end{array}\right)} =
 \djj {\fduvjj}^{\dagger}
{\left(\begin{array}{cc} {{\displaystyle\frac{\delta}{\delta u}}} \\
{\displaystyle{\frac{\delta}{\delta v}}} \end{array}\right)} H= \frac{1}{3}
{\left( \begin{array}{c} (j_{1x}+2j_{2x}+j_{1}^{2}-2j_{2}^{2}-2j_1 j_2 )_x \\
(-2j_{1x}-j_{2x}-2j_{1}^{2}+j_{2}^{2}-2j_1j_2)_x \end{array}\right)}\, ,
\ee
where we used ${\displaystyle\rhu}=\tilde{F}=0$,
${\displaystyle\rhv}=\tilde{C} =-\frac{1}{2}$ in the case of the Bsq equation.
The corresponding poisson brackets is defined by the hamiltonian operator of
mBsq equation.
Setting $J_1=\frac{\sqrt{3}}{2}(j_1+j_2), J_2=\frac{1}{2}(j_1-j_2)$,
we obtain
the poisson brackets in terms of two bosons $J_1$, $J_2$
that express the free field realization of $W_3$ algebra.
\newline\indent
We now find the Ur-Bsq equation.
Through the Fr\'{e}chet derivative and its formal adjoint of the transformation
eq.(\ref{17}),
the hamiltonian operator of Ur-Bsq equation which defines its
hamiltonian system and the poisson structure is related with the hamiltonian
operator of mBsq equation as eq.(\ref{16}).
The inverse of Fr\'{e}chet derivative
and its formal adjoint of the transformation eq.(\ref{17}) are easily obtained
by the definition of the Fr\'{e}chet derivative;
\bea \label{22}
\fdshjj\!\!\!\!&=&\!\!\!\!\left(\begin{array}{cc}
-2\pbi{\psi}_{x}\pbi & -\pbi{\psi}_{x}\pbi \\  -\pbi{\phi}_{x}\pbi &
2\pbi{\phi}_{x}\pbi \end{array}\right), \nonumber \\
{\fdshjj}^{\dagger}\!\!\!\!&=&\!\!\!\!\left(\begin{array} {cc}
-2\pbi{\psi}_{x}\pbi & -\pbi{\phi}_{x}\pbi \\  -\pbi{\psi}_{x}\pbi &
2\pbi{\phi}_{x}\pbi \end{array}\right) \, .
\eea
Using these differential operators and the hamiltonian operator of mBsq
equation, the hamiltonian operator of Ur-Bsq equation is
\be \label{23}
\dsh=\left(\begin{array}{cc} 2\pbi{\psi}_{x}\pbi{\psi}_{x}\pbi &
\pbi{\psi}_{x}\pbi{\phi}_{x}\pbi \\  -\pbi{\phi}_{x}\pbi{\psi}_{x}\pbi &
2\pbi{\phi}_{x}\pbi{\phi}_{x}\pbi \end{array}\right)\, .
\ee
This hamiltonian operator defines the poisson structure of Ur-Bsq equation;
\bea \label{24}
\left\{{\psi}(x), {\psi}(y)\right\}\!\!\!\!&=&\!\!\!\!
2\pbi{\psi}_{x}\pbi{\psi}_{x}\pbi\delta(x-y),\quad
\left\{{\psi}(x), {\phi}(y)\right\}=
\pbi{\psi}_{x}\pbi{\phi}_{x}\pbi\delta(x-y)\, , \nonumber \\
\left\{{\phi}(x), {\phi}(y)\right\}\!\!\!\!&=&\!\!\!\!
2\pbi{\phi}_{x}\pbi{\phi}_{x}\pbi\delta(x-y)\, ,
\eea
where we note that $\psi$ and $\phi$ separately satisfy subalgebras, each of
which is that of the Ur-KdV equation.
We can see clearly that the symplectic form associated with the
corresponding hamiltonian structure of the Ur-Bsq equation which is given by
an inverse of the poisson structure is local.
the Ur-Bsq equation is easily obtained, through the inverse Fr\'{e}chet
derivatives of $\psi, \phi$ with respect to $j_1, j_2$	and the
transformation eq.(\ref{17}), as follows:
\be \label{25}
\psi_t=
\frac{1}{3}\left(-{\psi}_{xx} -2{\phi}_{xx}{\phi}_{x}^{-1}{\psi}_{x}\right),
\quad
\phi_t=
\frac{1}{3}\left({\phi}_{xx}+2{\psi}_{xx}{\psi}_{x}^{-1}{\phi}_{x}\right)\, .
\ee
There are a relation between these equations and the Ur-Bsq equation given in
Ref.\cite{Sc}. That is, redefining of ${\tilde{\psi}}=\phi$,
${\tilde{\phi_x}}=\phi_x\psi$, we can recover the Ur-Bsq equation
given in Ref.\cite{Sc} of ${\tilde{\psi}}$, ${\tilde{\phi}}$.
Using an obvious form of Cole-Hopf map for $SL(N)$, $N>3$ cases, we expect to
get quite a simple form for Ur equations for these cases. Of course details
has to be worked out yet.
\newline\indent
The work presented here clarifies the relation for $SL(3)$ Ur-KdV equation
corresponding to $SL(3)$ KdV equation.
While the poisson structure of the usual KdV equation defined by its
hamiltonian
operator contains highly local differential operators, that of the Ur-KdV
equation is non-local operators.
Because the symplectic form is associated with the inverse of poisson brackets
in a classical mechanical system, the symplectic form of KdV system is
non-local and that of Ur-KdV system local. Therefore, from the hamiltonian
viewpoint, one say that the one flow takes place on a poisson manifold, the
other on a symplectic manifold, and then the
hamiltonian operators of two system (in the case of $SL(2)$, ($u$,$\dsu$),
($q,\dq$)) is called the antiplectic pair by Wilson\cite{Wi}.
It was noted\cite{Sc}
that the above observation gives a clue to look for a local action of KdV
system. In this paper, we found the Ur-KdV systems in the point of view what
we must have a transformation by which can be obtained the inverse Fr\'{e}chet
derivatives of the Miura map. In addition, this view point clarifies the
relation between the hamiltonian operators of KdV, mKdV and Ur-KdV equation.
\newline\indent
This work was supported in part by Ministry of Education, and by Korea Science
and Engineering Foundation through SNU/CTP.

%

\end{document}